\documentclass[graybox]{svmult}

\usepackage{mathptmx}       
\usepackage{helvet}         
\usepackage{courier}        
\usepackage{type1cm}        
\usepackage{makeidx}         
\usepackage{graphicx}        
\usepackage{multicol}        
\usepackage[bottom]{footmisc}

\usepackage[numbers]{natbib}

\usepackage{url}

\usepackage{amsmath,amssymb} 

\makeindex             

\newcommand{\dg}{\mbox{$^{\circ}$}}

\begin{document}

\title*{Magnetic Fields in the Milky Way}
\author{Marijke Haverkorn}
\institute{Marijke Haverkorn \at Department of Astrophysics/IMAPP,
Radboud University Nijmegen, P.O. Box 9010, 6500 GL Nijmegen, The
Netherlands;  Leiden Observatory, Leiden University, P.O. Box 9513,
2300 RA Leiden, The Netherlands \email{m.haverkorn@astro.ru.nl}}

\maketitle

\abstract{
  This chapter presents a review of observational studies to determine
  the magnetic field in the Milky Way, both in the disk and in the
  halo, focused on recent developments and on magnetic fields in the
  diffuse interstellar medium. I discuss some terminology which is
  confusingly or inconsistently used and try to summarize current
  status of our knowledge on magnetic field configurations and
  strengths in the Milky Way. Although many open questions still
  exist, more and more conclusions can be drawn on the large-scale and
small-scale components of the Galactic magnetic field. The chapter is
concluded with a brief outlook to observational projects in the near future.}

\section{Introduction}

The Milky Way is a dynamic environment, much of which (partially)
consists of plasma: stars, jets, objects such as H~{\sc ii} regions or
supernova remnants, and the general interstellar medium (ISM). No
wonder that magnetic fields are ubiquitous throughout the Galaxy, in
almost all astrophysical objects from strong fields in pulsar
atmospheres to weak fields on scales of many kiloparsecs, threading
the whole Galaxy. The importance of these magnetic fields is manifold:
in the energy balance of the Milky Way, transport of angular momentum,
acceleration and propagation of charged particles, gas dynamics,
etc. All interstellar matter but the densest, coldest clouds is
sufficiently ionized (even with an ionization degree of only $10^{-4}
- 10^{-3}$) for the neutral gas component to remain coupled to the
ionized gas, and therefore be efficiently frozen into the magnetic
field \citep{f01}. Equipartition of magnetic and turbulent gas density
\cite{heha12} indicates that dynamical feedback of the magnetic field
on the gas plays an important role.

Fully characterizing the strength, direction, and structure of the
extended Galactic magnetic field threading the entire Milky Way is an
extremely daunting task. This field can be regarded as a combination
of a large-scale field threading the Galaxy, probably maintained by
the Galactic dynamo, and a small-scale field. The small-scale field is
caused by and interacts with interstellar turbulence, supernova
explosions and remnants and other shock waves, and is altered by gas
dynamics, magnetic reconnection, turbulence effects etc. In addition,
the available observational methods detect either one component of the
magnetic field (strength or direction, parallel or perpendicular to
the line of sight) and/or in one particular tracer (ionized gas, dense
cold gas, dense dust, diffuse dust).
Lastly, some of the difficulty of determining
the Galactic magnetic field stems from our vantage point inside the
Milky Way. Creating a three-dimensional picture from mostly
two-dimensional tracers necessitates many assumptions about the
magnetic field, as well as about the thermal and cosmic ray electron
distributions, and about the (local) interstellar objects and
processes influencing these.

Despite the difficulties, attempts to detect and determine the
Galactic magnetic field have been many in recent (and not so recent)
years. This is not only because magnetic fields influence so many
physical processes in the ISM, but also because of its importance to
other fields in astronomy and astrophysics. For instance, the Cosmic
Microwave Background (CMB) community has shown a keen interest in the
Galactic magnetic field, since it produces Galactic polarized
synchrotron emission which acts as a strong foreground for CMB
polarization. Also, astroparticle physicists studying sources and
propagation of Galactic and extragalactic cosmic rays profit from
detailed magnetic field models, which predict distributions of arrival
directions of (high energy) cosmic rays. In addition, high-precision
cosmological studies of the Epoch of Reionization need a detailed
understanding of Galactic polarization to be able to understand and
subtract any polarization leaking into their extremely sensitive
measurements of highly redshifted H~{\sc i}.

It is not possible to cover all observations of magnetic fields in the
Milky Way in this review. Fortunately, I can refer to a number of
complementing reviews. For observations of magnetic fields in dense
clouds and their relation to star formation, see 
various chapters in this Volume. For a historical review on magnetic
field observations in the Milky Way, see \cite{w05} or \cite{wb10},
and \cite{f09} provides an excellent treatise on magnetic fields in
the Galactic Center. I refer to e.g.\ \cite{f07} for a review on the
very local ISM, including magnetic fields, and to \cite{n12} for a
recent review on Galactic magnetic fields from Faraday rotation of
pulsars and extragalactic sources. \cite{hh12} published a recent
review about magnetic fields in galactic haloes. I will focus here on
work mostly in the last decade. For earlier reviews, see \cite{h01,
  w05, n12, heha12}.

This chapter starts with a brief description of some terminology used in
literature in Section~\ref{s:term}. Sections~\ref{s:disk},
\ref{s:halo} and~\ref{s:diskhalo} describe current knowledge of
magnetic field observations in the Galactic disk, in the Galactic halo
and in the combined disk-halo system, respectively. A short summary
and conclusions are stated in Section~\ref{s:conc}, and finally,
Section~\ref{s:future} describes some recent, progressing and future
observational projects which are important for the investigation of
Galactic magnetic fields.

\section{Terminology}
\label{s:term}

\subsection{Large-scale vs small-scale fields}
\label{s:lsvss}

The description of large-scale and small-scale galactic magnetic
fields in the literature is often confusing, with different authors
using different terminology for the same magnetic field configurations
or the same words for different field structure. Here I give an
overview over these different magnetic field configurations and ways
to describe them.

Traditionally, galactic magnetic fields have been divided up in
small-scale and large-scale fields. The term ``large-scale'' fields
(also called regular, uniform or coherent fields) indicates the
component of magnetic field that is coherent on length scales of the
order of a galaxy, usually assumed to follow the spiral arms or to be
ring-shaped. ``Small-scale magnetic fields'' (also called random,
tangled, or turbulent) describe the magnetic field component connected
to the turbulent ISM. The small-scale field is usually
simply assumed to follow a power law with a certain outer scale,
where energy is injected, which then cascades down to smaller
turbulent scales until energy dissipates at the dissipation
scale. Small-scale magnetic field fluctuations connected to discrete
objects such as H~{\sc ii} regions or supernova remnants warrant their
own review paper and are usually treated separately from the
``Galactic magnetic field'', although interaction between these fields
and the general Galactic magnetic field is of course pervasive.

\begin{figure}
\centerline{\includegraphics[width=0.8\textwidth]{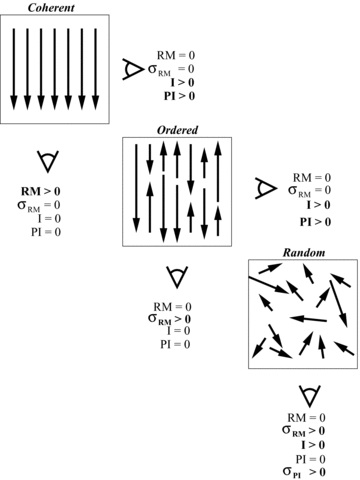}}
\caption{Sketch illustrating the three components of galactic magnetic
  fields. For consistency in the literature, the component labeled
  ``ordered'' here should be called ``ordered random'' and the
  component labeled ``random'' should be ``isotropic random''. The
  three observables for these magnetic fields are total intensity $I$,
  polarized intensity $PI$ and rotation measure RM. Image reproduced
  from \cite{jlb10}.}
\label{f:components}
\end{figure}

However, lately, a third component of the magnetic field starts to be
included in Galactic magnetic field studies, as it has for some time
in magnetic field studies of external galaxies (see R. Beck's Chapter
in this Volume). This component is a field of which the direction
varies on small scales, but the orientation does not. Such a field can
arise when a turbulent field structure is compressed into a
two-dimensional structure by e.g.\ supernova remnant shocks, spiral
arm density waves, or galactic shear. This field component is often
referred to as anisotropic random, but is also called ordered random
\citep{b07} or striated \citep{jf12a}\footnote{Note that \citep{jlb10}
  refer to this component as 'ordered'}. A clear explanation of these
components is given in Fig~\ref{f:components}, reproduced from
\cite{jlb10}. The cartoons illustrate the morphology of the three
components and indicate the differences between the tracers total
intensity $I$, polarized intensity $PI$ and rotation measures RM for
different lines of sight towards these three components. Combination
of these tracers makes it possible to distinguish between the three
field components.

As Fig~\ref{f:components} shows, studies using RMs alone cannot
distinguish between ordered random and isotropic random field
components, which are often grouped together in a ``random''
field. Similarly, investigations using synchrotron emission cannot
distinguish between coherent and ordered random field, due to which
these two components are often assembled into one ``ordered''
component.

\subsection{Configurations of large-scale Galactic magnetic fields}

\begin{figure}
\centerline{\includegraphics[width=\textwidth]{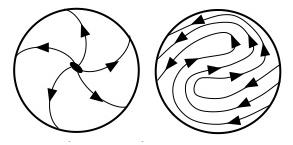}}
\caption{Magnetic field configurations of the disk field: bird's-eye view of a
  galaxy with axisymmetric (left) vs.\ bisymmetric (right) spiral
  magnetic field lines in the galactic disk. Image adapted from
  \cite{b10} and \cite{zh97}.}
\label{f:conf_assbss}
\end{figure}
\begin{figure}
\includegraphics[width=\textwidth]{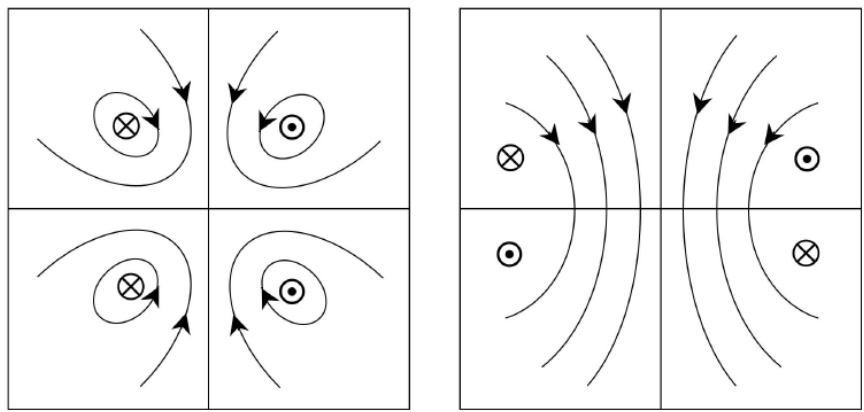}
\caption{Magnetic field configurations of the halo field: edge-on view
  of a galaxy with symmetric, even field configuration (left) or
  anti-symmetric, odd field configuration. Image reproduced
  from \cite{hh12}.}
\label{f:conf_evenodd}
\end{figure}

A number of fairly simple configurations have been explored for the
coherent magnetic field in the Milky Way. These configurations are
based on rotational symmetry around the Galactic Center, and on mirror
symmetry with respect to the Galactic plane.

The two simplest disk configurations are an {\em
  axisymmetric}\footnote{sometimes called {\em disymmetric}
  \citep{jfw09}.} and {\em bisymmetric} spiral structure, which denote
magnetic fields oriented along the spiral arms. In the axisymmetric
situation, magnetic field lines all point inwards or
outwards. Bisymmetric fields are antisymmetric with respect to the
spin axis of the galaxy. Therefore, the bisymmetric situation includes
field reversals in the azimuthal direction (see
Fig~\ref{f:conf_assbss}). Axisymmetric fields are denoted by azimuthal
mode $m=0$, bisymmetric fields are $m=1$. Higher azimuthal modes $m$
or a mix of modes may be present, e.g.\ the $m=2$ or {\em
  quadri-symmetric} mode.

The field can also be described in terms of its symmetry with respect
to the Galactic plane. A {\em symmetric or even-parity} field has a
mirrored magnetic field configuration (see
Fig~\ref{f:conf_evenodd}). Note that this indicates a reversal of the
vertical magnetic field direction across the Galactic plane, and that
the toroidal component of the magnetic field points in the same
direction above and below the plane. An {\em anti-symmetric or
  odd-parity} field has field lines that run through the Galactic
plane, and toroidal fields that reverse direction above and below the
plane.

This symmetric and anti-symmetric mirror symmetry is often denoted
with $S$ and $A$, respectively. This classification is followed by a
number which gives the azimuthal mode number $m$. So, e.g., an $A0$
field configuration has an axisymmetric field, the horizontal
component of which is directed in opposite directions above and below
the Galactic plane.

A slightly more complex field configuration is the Disk-Even-Halo-Odd
(DEHO) configuration, consisting of two independent field components
for the Galactic disk and halo; as the name indicates, the vertical
symmetry of this magnetic field configuration is even in the Galactic
disk, but odd in the halo. I mention this morphology here, since it is
preferred in several observational studies of all-sky magnetic field
configurations, discussed in Section~\ref{s:diskhalo}.

\subsection{Pitch angle definition}
\label{s:pitch}

The pitch angle of a spiral magnetic field is defined as 
\begin{equation}
p = \tan^{-1}\frac{B_r}{B_{\phi}}
\end{equation}
where $B_r$ is the radial component of the magnetic field and
$B_{\phi}$ its azimuthal component. For a trailing spiral, $B_r$ and
$B_{\phi}$ have opposite signs, so that the pitch angle is negative,
also described in the literature as ``radially inward''. Note that
\cite{phk07} use a deviating definition of their angle $\psi_0$ as
$\psi_0 = \tan^{-1}B_{\phi}/B_r$.

\section{Magnetic fields in the Galactic disk}
\label{s:disk}

\subsection{Large-scale magnetic field strength}

The strength of the local large-scale magnetic field as obtained from
Faraday rotation of pulsars and extragalactic sources is typically
around $1.5 - 2~\mu$G. These estimates mostly result from RM and
Dispersion Measure (DM) data from pulsars \citep{m74, hml06}, from
wavelet analysis \citep{fss01} or fitting RM data to large-scale
models of Galactic magnetic fields (as discussed extensively below).

The total field strength in the Solar neighborhood is estimated to be
around $6~\mu$G, from observed synchrotron emissivities and assumed
equipartition between cosmic rays and magnetic fields
\citep{smr00,b01}. This is in agreement with magnetic field strength
estimates in Galactic H~{\sc i} regions from Zeeman splitting ($B
\approx 2-10~\mu$G, \cite{cwh10}).

Towards the Galactic center, the magnetic field strength
increases. Estimates from synchrotron emission give a total field
strength of about $10~\mu$G at a Galactocentric radius of 3~kpc
\citep{b01}, the pulsar study by \cite{hml02} concludes a regular
field strength of $4.4\pm0.9~\mu$G in the Norma arm, and also
large-scale magnetic field modeling generally finds stronger total
magnetic field strength towards the Galactic center
\citep[e.g.,][]{bhg07, smr00, nk10}. The extensive study by \cite{h95},
using various tracers, concludes that $B_{tot} \approx 7.6 -
11.2~\mu$G at a Galactocentric radius of 4~kpc.

The magnetic field strength is independent of density for low
densities in the diffuse ISM ($n \lesssim 300$~cm$^{-3}$), indicating
infall along magnetic field lines \citep{cwh10}. Only for dense clouds
and molecular clouds, magnetic field strengths increase  roughly as
the square root of density.

\subsection{Large-scale magnetic field structure}

The configuration of the large-scale magnetic field in the Milky Way
disk is still a matter of hot debate. Some features meet with
reasonable or total agreement: the magnetic fields seem to roughly
follow the spiral arms, which is in agreement with all external spiral
galaxies observed \citep{b01}, and even ring galaxies \citep{cb08}
(however, see \cite{v08}). This conclusion is drawn not only from
polarized radio synchrotron and Faraday rotation measurements, but is
also supported by starlight polarization measurements \citep{h96,
  nht10}, submm dust polarization \citep{bmd11} and Zeeman splitting
observations in hydroxyl masers \citep{gmc12}. Even young H~{\sc ii}
regions \citep{pc12} and molecular clouds \citep{hz07} seem to have
magnetic fields aligned with a large-scale field along the Galactic
plane .

Also, one large-scale reversal of the magnetic field near the Sun
towards the Galactic Center has been known for decades \citep{tn80,
  sk80} and is confirmed by the rotation measure studies discussed
here, but also by magnetic field directions in massive star-forming
regions as probed by Zeeman splitting of OH masers
\citep{fra03}. However, the exact number and location of large-scale
reversals, pitch angles, characteristics of the turbulent magnetic
field as a function of location and properties of the magnetic field
close to the Galactic Center are still under discussion.

The past decade has seen a surge in studies using ad-hoc Galactic
magnetic field models such as axisymmetric, bisymmetric or ring-shaped
magnetic fields to fit to observational data. The goal is to determine
free fit parameters such as magnetic field configuration, pitch angle
and strength. Table~\ref{t:bfldmodels} shows a brief and necessarily
incomplete summary of these models and some of their properties, as an
attempt to make the differences between these models insightful, and
to draw conclusions from this large body of work by many authors.
Many of these models contain complexities that cannot all be captured
in a simple table, e.g.\ models include radially declining magnetic
field strengths or use different ways to incorporate large-scale
magnetic field reversals.  Also, the models use various models of
thermal and/or cosmic ray electron densities, which we do not discuss
here at all. The range of conclusions in these papers is much wider
than noted in the table; here we focus on modeling results about the
magnetic field strength and structure only.

It is highly non-trivial to compare the results from these models
since they are so heterogeneous: most models use different input
configurations for magnetic fields, thermal electron density and
cosmic ray density, and use the various magnetic field parameters as
either input or output parameters. However, some consensus seems to
appear: most models tend to favor axisymmetric magnetic field models
with one reversal just inside the Solar circle \citep{bhg07, srw08,
  jfw09, jlb10}. These best fit configurations (sometimes with some
embellishments) have been taken as fixed input in subsequent papers,
in order to determine e.g.\ out-of-plane magnetic fields
\citep{jf12a}, or the pitch angle and synchrotron spectral index
\citep{fma11}.  However, careful analysis of pulsar RMs by
\cite{mfh08} proved that none of the three widely used magnetic field
models (axisymmetric, bisymmetric, ring) are consistent with the
data. These authors conclude that the magnetic field of the Milky Way
must be more complex than one simple dynamo mode, possibly a
combination of modes, as observed in some external galaxies (see
R. Beck's Chapter, this Volume).

One notable difference in results can be seen in models based mostly
on pulsars and models based mostly on extragalactic sources. RMs of
extragalactic sources average magnetic field and density fluctuations
over the complete line of sight through the whole Galaxy. Pulsar RMs
only probe the line of sight to the individual pulsar, or even the
path length between two pulsars in close projected proximity on the
sky, which is a shorter distance and much more variable over small
coordinate differences. In addition, RMs from extragalactic sources
tend to be averaged over some region in the sky in order to diminish
contributions from their intrinsic RM and from the turbulent Galactic
ISM. Therefore, RMs measured from pulsars tend to show much more
influence of the small-scale magnetic field component.  Good examples
of this are presented in \cite{hml06}, who used RMs from
pulsars. They did not use any model but constructed a magnetic field
configuration by looking at sign reversals of pulsar RMs in arms or
interarms. Their data confirmed a counter-clockwise field in the
Carina-Sagittarius spiral arm and suggest a counter-clockwise field in
the Perseus. They find an abundance of small-scale structure in RM
sign, which they interpret as clockwise magnetic fields in the
interarm regions and counter-clockwise magnetic fields in the spiral
arms, indicating large-scale magnetic field reversals at every
arm-interarm boundary. Magnetic field modeling by \cite{nk10} confirm
reversals at every arm-interarm boundary, but find results at
$>3\sigma$ only for the Crux and Norma arms and the interarm region in
between. 

\begin{figure}[!h]
\includegraphics[width=\textwidth]{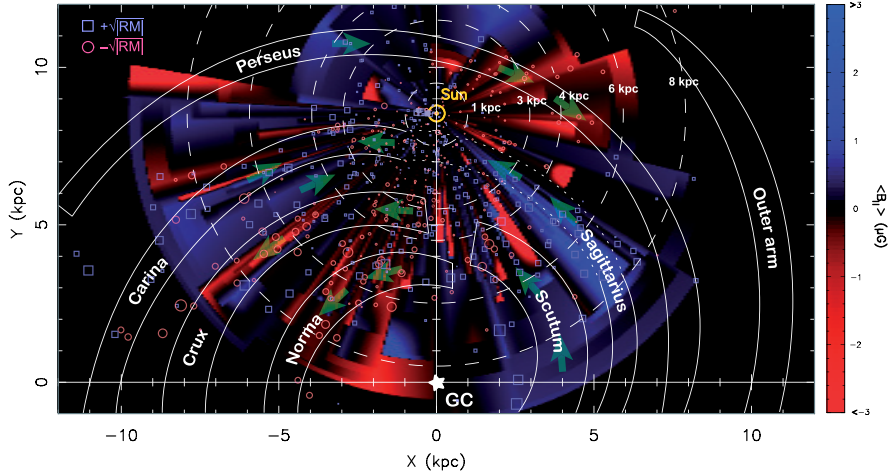}
\caption{ Bird's-eye view of the Milky Way, where the blue squares (red
  circles) denote the location of a pulsar with a positive (negative)
  RM. The size of the symbol is proportional to the square root of
  RM. Magnetic field strengths derived and interpolated from these
  pulsar data are given in red-blue color scale (where $B > 3~\mu$G is
  saturated). The green arrows give the predominant direction of 
  parallel magnetic field in a certain region. Image reproduced from
  \cite{n12}.}
\label{f:noutsos12_fig1}
\end{figure}

\begin{figure}[!h]
\includegraphics[width=0.45\textwidth]{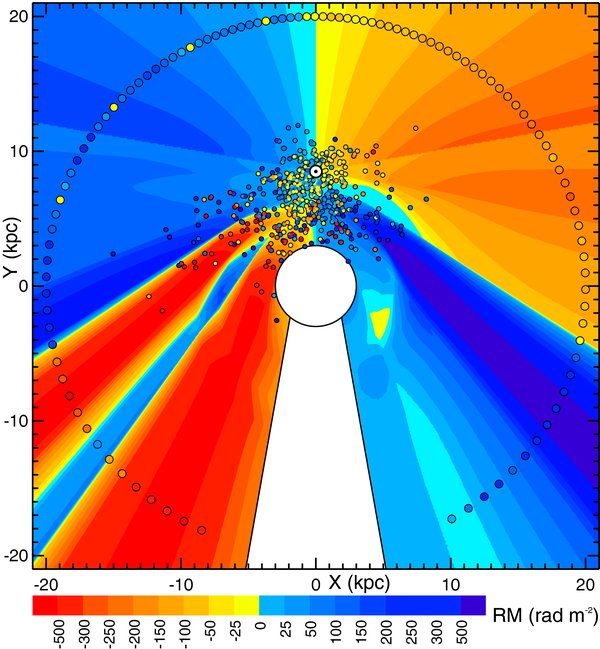}
\includegraphics[width=0.55\textwidth]{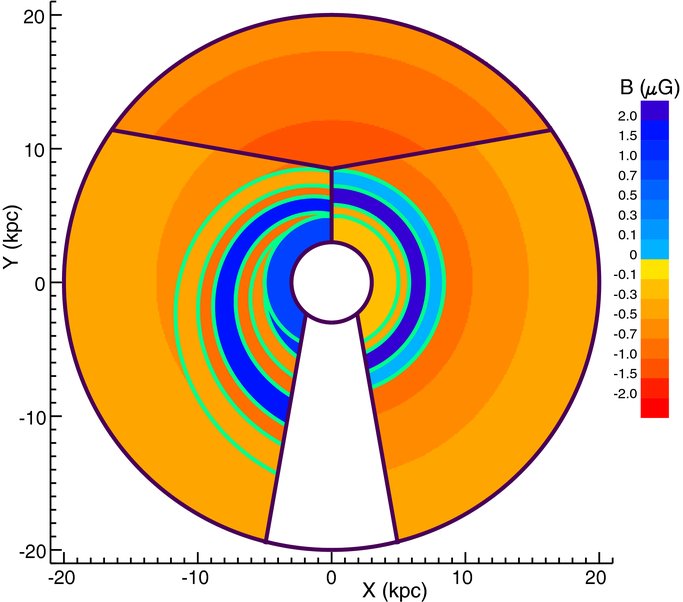}
\caption{{\em Left:} Bird's-eye view of the Milky Way, where the
  Galactic Center is at $(X,Y) = (0,0)$ and the Sun is at $(X,Y) =
  (0,8.5)$~kpc. The small circles within the Galaxy denote observed
  pulsar RMs, large circles around the Galaxy show observed
  extragalactic source RMs in the Galactic plane, boxcar-averaged over
  9\dg~in longitude with a step size of 3\dg. The background color
  scale presents predicted RMs at each location according to the model
  in the right hand figure. {\em Right:} Model of Galactic magnetic
  field in which the Galaxy is divided up into three regions. Color
  denotes magnetic field strength. Outer Galaxy: logarithmic spiral
  with $p = -11.5$\dg; fourth Galactic quadrant: model from
  \cite{bhg07}; first quadrant: ASS+RING model from \cite{srw08}.
  Image reproduced from \cite{vbs11}.}
\label{f:vaneck12_1012}
\end{figure}

Fig~\ref{f:noutsos12_fig1}, reproduced from \cite{n12},
nicely illustrates the intermediate-scale structure in RMs from
pulsars, which are interpreted in the literature as reversals along
spiral arm directions \citep{hml06} or as intermediate-scale
fluctuations in the field \citep{n12}.
As an example of the difference with modeling results including
extragalactic sources, I mention \cite{vbs11}, who analyzed the Milky
Way's magnetic field RM data of pulsars and extragalactic sources
combined. They divide up the Galactic disk in three separate longitude
ranges and concluded that there is no simple configuration which fits
the whole Galactic plane sufficiently well, see
Fig.~\ref{f:vaneck12_1012}. They also conclude that not more than one
large-scale field reversal is needed to explain the data.

One way to decrease the influence of small-scale structure on pulsar
RM measurements is by averaging these data as well before analysis of
the structure. This can be done e.g.\ by wavelet analysis
\citep{sfs02}, using pulsar RMs. Using this method, \citep{fss01} only
obtained reliable results a few kpc from the Sun due to sparsity of
data beyond. However, these authors found evidence for one magnetic
field reversal at a distance of $0.6-1$~kpc towards the Galactic center,
and an other reversal between the Perseus and (local) Orion arm, in
agreement with some earlier studies \citep{rk89, ccs92}. However,
\cite{mwk03} show that the anomalous RMs interpreted as a large-scale
reversal towards the Perseus arm can be explained by anomalous RMs due
to the influence of H~{\sc ii} regions along the line of sight.  

There is some evidence to suggest that the one well-determined
large-scale reversal in the disk magnetic field does not follow the
Sagittarius-Carina arm exactly, but slices through it \citep[][see
also Fig.~\ref{f:vaneck12_1012}]{bhg07, vbs11}, a phenomenon that has
been seen in the nearby spiral galaxy M51 as well \citep{fbs11,hbe09}.
In addition, the magnetic field towards the outer Galaxy $l\sim
180\dg$ may be closer to circular rather than spiral \citep{rb10,
  vbs11}.

Also, many studies provide evidence for a dominant even symmetry of
the local regular field in the disk with respect to the Galactic plane
\citep{fss01,jfw09,ptk11,mmg12}.

\begin{table}
  \caption{Table summarizing models comparing Milky Way magnetic field
    configurations to various observational tracers. Note the
    incredible range of possible data, models, fixed parameters and
    output parameters.  Column~1 gives the reference to 
    the paper, column~2 details the used tracers and column~3 notes
    whether the model pertains to the Galactic disk, the halo, or both
    ('all'). Column~4 summarizes the ad-hoc models used for each
    paper, and Column~6 (some of) the main results. Column~7 gives the
    pitch angle of the {\em disk} field, where 'IN' is added if this pitch angle was a fixed
    input value. Symbols are as defined in the text. 
  }
\label{t:bfldmodels}
\begin{tabular}{p{0.5cm}p{2.7cm}p{0.7cm}p{2.5cm}p{3.6cm}p{1.0cm}}
\hline\noalign{\smallskip}
ref & TRACER$^a$ & D/H & MODELS$^b$ & MODEL RESULTS & $p$\\
\hline
\cite{bhg07} & 149 EGS RMs     & Q4$^c$ disk & spiral & one reversal & $-11.5$\dg\\ 
                      & 120 pulsar RMs &  &&&\\
\hline 
\cite{fma11} & WMAP5 $I$ 23GHz; & all & modified log spiral & $B_z =
0.4~\mu$G & $-30$\dg\\
                      & ARCHEOPS 353GHz                & & $B_z$ + $B_{ran}$  &&\\
                      &$I$ 408MHz              & &  &&\\
\hline 
\cite{jlb10} & $I$ 408MHz & disk & ASS, log spiral, &
$B_{reg}:B_{ran}:B_{ani} = 1:5:4$ & $-11.5$\dg~IN\\
                    & WMAP $P$ 23GHz & & $B_{ran}$, compression&Field
                    config as in model 1&\\
                    & 269 EGS RMs &&&&\\
\hline 
\cite{jfw09} & WMAP5 $PI$ 23GHz & disk & BSS/ASS -S/-A, & no good models, &
$+35$\dg\\
                     & 1433 EGS RMs & & ring, lit.\ models  &
                     disk and halo separate &\\
\hline 
\cite{jf12a} & WMAP7 $PI$ 23GHz & all & spiral, $B_{ran}$, $B_{ani}$,
&one reversal $B_{ani} = 1.7B_{reg}$, &$-11.5$\dg~IN\\
                  &  $\gtrsim$37000 EGS RMs & &$B_z$ & $B_z= 4.6~\mu$G at GC$^d$ &\\
\hline 
\cite{mfh08} & 482 pulsar RMs & disk & ASS, BSS, ring & no good models,
slight preference for ASS & \\
\hline 
\cite{myl08} & $I$ 408MHz & halo & BSS, $B_{ran}$ & $B_{ran} = 0.57 B_{reg}$&$-8.5$\dg\\
                      & WMAP $PI$ 23GHz & &&&\\
\hline 
\cite{nk10} & 133 pulsar RMs& Q4$^d$& log spirals &	QSS/many
reversals preferred &\\	
   & 107 EGS RMs& disk&  &&	\\	
\hline 
\cite{phk07}& WMAP3 $PI$ 23GHz & halo &	log spirals, $B_z$
&$B_z$ at $25^{\circ}$ tilt & $-55$\dg~$^d$\\
\hline 
\cite{ptk11}& $\gtrsim$37000 EGS RMs & all & ASS, BSS, ring & ASS
best in disk; odd in halo&$-5$\dg\\
    &              &  &&&\\
\hline 
\cite{rrb10}& WMAP5 $PI$ 23GHz & halo & ASS, BSS, ring, 
&  ASS preferred,  $B_z=1~/mu$G& $-24$\dg$^e$\\
 &&&bi-toroidal, $B_z$ &&\\
\hline 
\cite{srw08}& $I$ 408MHz & all & ASS, BSS, ring & ASS best in disk,
odd in halo & $-12$\dg~IN\\
                     & WMAP $PI$ 23GHz & &&&\\
                     & $I + PI$ 1.4GHz &&&&\\
\hline 
\cite{v05}& 354 pulsar RMs & disk & rings with $p$  & one reversal only
& $-12$\dg~IN\\
\hline 
\cite{vbs11}& 1373 EGS RMs & disk &ASS, BSS, ring& no single
model & 0\dg or \\
                      & 557 pulsar RMs	&& combinations& for complete Galaxy &$-11.5$\dg~IN\\
\noalign{\smallskip}\hline\noalign{\smallskip}
\end{tabular}
$^a$ $I$ = total intensity; $PI$ = polarized intensity; EGS =
extragalactic sources; WMAP$i$ = Wilkinson Microwave Anisotropy Probe
data over $i$ years. \\
$^b$ ASS = axisymmetric spiral; BSS = bisymmetric spiral; QSS
= quadrisymmetric spiral;  -A/-S = (anti-)symmetric with respect to
Galactic plane. \\
$^c$ Q$i$ = $i$th quadrant of the Milky Way; GC = Galactic Center.\\
$^d$ taking into account their deviating definition of pitch angle,
see Section~\ref{s:pitch}.\\
$^e$ actually given as $p=+24\dg$ in the paper, but with the opposite
definition of azimuth direction.
\end{table}

\subsection{The pitch angle of the magnetic spiral arms}

Estimates of the pitch angle of the Milky Way's magnetic spiral arms
are widely varying, depending on the tracer used to determine this
angle. \cite{v95,v02} collected pitch angle estimates from 1980 to
2001, obtained from H~{\sc i} and H~{\sc ii} gas, pulsars, dust, CO,
rotation measures and O stars, which vary from $-5^{\circ}$ to
$-21^{\circ}$. His weighted average is $p = -12^{\circ} \pm
1^{\circ}$. Polarized starlight indicates a pitch angle of $p =
-7.2\dg \pm 4.1\dg$ \citep{h96}, consistent with the other estimates.

The magnetic field models in Table~\ref{t:bfldmodels} based on RM data
give pitch angles in the range $p \sim -5\dg$ to$-15\dg$, while models
fitting to high-frequency (mostly WMAP) polarized synchrotron emission
data tend to find much higher pitch angles $p\sim -25\dg$
to~$-30^{\circ}$ (except for \cite{myl08}, who find $p =
-8.5$\dg~based on the WMAP degree of polarization, modeling a BSS
spiral field and adding a turbulent component to match the observed
depolarization). As a notable exception, \cite{jfw09} finds a high
(and oppositely directed!)  pitch angle $p = +35^{\circ}$, based on
RMs, but warns that this result is ``highly model
dependent''. Indeed, they describe that fixing the pitch angle at $p =
-12^{\circ}$ in their best-fit model only decreases the fitting
quality slightly. 

Using the straightforward method of comparing longitude-dependences of
RMs of extragalactic sources in the first and fourth quadrants,
\cite{kn11} conclude that the Milky Way has a bisymmetric structure
towards the inner disk and an axisymmetric pattern towards the outer
disk with an ``inward spiral pitch angle'' of $5.5^{\circ}\pm
1^{\circ}$. However, their pitch angle calculation assumes that the
sign changes in RM in the first and fourth quadrant are due to the
same spiral arm, which would actually indicate a pitch angle of $p =
+5.5^{\circ}$, i.e.\ a trailing instead of a leading spiral. The
longitudes of the changes in RM sign in the first and fourth quadrants
are more plausibly due to the Local Arm in the first quadrant and the
Carina arm in the fourth, in which case their pitch angle estimate is
based on an incorrect assumption.

\subsection{Turbulent magnetic fields in the disk}

The strength of the random, turbulent magnetic field component can be
estimated from RM fluctuations, combined with an estimate of thermal
electron density\footnote{Since RM fluctuations and synchrotron
  depolarization trace the isotropic random
field and the parallel component of the ordered random field (see
Fig.~\ref{f:components}), the strengths cited are a combination of
these two components.}. \cite{gdm01} performed this analysis in a small
region in the Galactic plane in the 4th quadrant and found a random
field strength of $B_{ran} \gtrsim 1.3~\mu$G. Large-scale magnetic
field models that include the turbulent magnetic field as a free
parameter find $B_{ran} \sim 3-4~\mu$G \citep[e.g.,][]{jlb10,
  srw08}. However, the estimates for the regular and total magnetic
field strengths would suggest a slightly larger value for the
turbulent component, i.e.\ $B_{ran} = \sqrt{B_{tot}^2 - B_{reg}^2}
\approx 5.5~\mu$G.

Under the assumption of a Faraday screen, \cite{skd07a} find that
$B_{ran}/B_{reg,\perp} \lesssim 2$ in a relatively small field of view
at the anti-center at Galactic latitude $b=20\dg$, based on
synchrotron depolarization. In two other relatively small fields out
of the Galactic plane, \cite{hkb04} find $B_{ran} \approx 1-3~\mu$G
and an unusually low ratio of random to regular magnetic field
components $B_{ran}/B_{reg} = 0.7\pm 0.5$. However, both these regions
are located in the extended, high-polarization Fan region, which is
thought to have a higher contribution of the regular magnetic field
than the average ISM \citep{wlr06}. This unusually low value of
magnetic field ratio agrees with the higher-frequency synchrotron
polarization study of \cite{myl08}, which however includes the
anisotropic random field component in their regular field strength
calculation, plausibly explaining the low random-to-regular magnetic
field ratio.

The power spectrum of magnetic field is difficult to measure directly,
but information about the field can be derived from power spectra or
structure functions of RM, assuming some distribution of thermal
electron density fluctuations. RM fluctuations follow a power law,
although the slope tends to be flatter than Kolmogorov \citep{sc86,
  ccs92}. Early studies covering large parts of the sky conclude that
the outer scale of the turbulent magnetic field, connected to the
energy injection scale of the dominant source, is $\sim 100$~pc
\citep{ls90,os93}. However, when distinguishing spiral arms and
interarm regions explicitly, the turbulent outer scale in spiral arms
seems to be much smaller, only a few parsecs
\citep{hgm04,hgb06,hbg08}. Small outer scale estimates like this are
also found from anisotropies in TeV cosmic ray nuclei \citep{mdd10},
and from analysis of fluctuations in radio synchrotron emission in the
Fan region \citep{iho13}.

Finally, there is evidence for an anti-correlation between small-scale
magnetic field structure and density, at least in the denser ISM:
denser components display more disordered magnetic field structure in
submm BICEP data in the Galactic plane \citep{bmd11}.

\section{Magnetic fields in the Galactic Halo}
\label{s:halo}

The strength of the magnetic field in the Galactic halo\footnote{Two
  separate definitions of the Galactic halo with respect to the thick
  disk cause some confusion: The Galactic halo is regularly referred
  to as the region above the thick disk. However, in a second common
  use of the term Galactic halo it is equal to the thick disk. We use
  here the second definition, where the halo is equal to the thick
  disk.} is estimated to be between $2-12~\mu$G from the best-fit
models in Table~\ref{t:bfldmodels}.
The field in the halo is thought to be fairly uniform: the average
line of sight component of the magnetic field at high latitudes (where
the $\sin(b)$ dependence has been taken into account) has a standard
deviation $\sigma_B \lesssim 0.4~\mu$G \citep{s10}.  Using
equipartition arguments, \cite{mgh10} derived $B_{ran} \approx 1~\mu$G
in the halo, indeed smaller than in the disk.

The scale height derived from synchrotron emissivity under the
assumption of equipartition between cosmic rays and magnetic fields is
about $5-6$~kpc \citep{c05}. Using hydrostatic balance, including
kinetic, magnetic and cosmic-ray pressures, \cite{bc90} find an
almost linearly decreasing field strength from about $5~\mu$G in the
plane to $1-2.5~\mu$G at 3~kpc height above the disk. \cite{hq94} used
pulsar RMs to derive a magnetic field scale height of 1.5~kpc - the
discrepancy with earlier estimates may be due to the fact that pulsar
RMs only sample the large-scale, regular component of the field while
equipartition estimates also take into account the turbulent
component.

The northern and southern hemisphere have different properties. The RM
variance is a factor of ~2 higher toward the South Galactic Pole than
toward the North Galactic Pole \citep{sts11}. RM data also show a
north-south asymmetry in RMs \citep{fss01}, emphasized by
\cite{mmg12}, who studied extragalactic source RMs in two distinct
parts of the sky towards the outer Galaxy ($100\dg < l < 117\dg$ and
$|b| > 15\dg$). They concluded that the observations cannot be
reproduced by symmetric exponential or double-toroidal Galactic halo
fields as used in the literature. They find a higher halo field
strength in the south ($B_H\approx 7~\mu$G) than in the north
($B_H\approx 2~\mu$G), and suggest that magnetic spiral arms might
exist in the halo as well.

A large-scale vertical magnetic field at the Solar radius is small, if
it exists at all. \cite{mgh10} find from extragalactic source RMs
towards the northern and southern Galactic pole at $|b| > 70^{\circ}$
that there is no evidence for a large-scale vertical magnetic field
component at the Solar radius in the northern hemisphere, while
$\langle B_z\rangle \approx 0.3~\mu$G in the south.  This is not
necessarily due to an asymmetry in the large-scale vertical field, but
can be due to differences in nearby structure in the two
hemispheres. \cite{tss09} evaluated the vertical magnetic field from
RMs from all NVSS\footnote{NRAO VLA Sky Survey \citep{ccg98}.}
sources. Their conclusion that $\langle B_z\rangle = 0.3\pm 0.03~\mu$G
agrees with \cite{mgh10} in the southern hemisphere, but they also
find a small vertical magnetic field of $\langle B_z\rangle =
-0.14\pm0.02~\mu$G in the northern hemisphere. \cite{mgh10} attribute
this difference to the North Polar Spur, which they removed from their
data and \cite{tss09} did not. \cite{hq94} found a vertical magnetic
field $B_z = 0.2-0.3~\mu$G from south to north; however, they forced
the direction of the field to be south-north or north-south and only
fitted the field strength. Therefore, it is not possible to say
whether their data would agree with the above conclusions.  Small and
varying vertical magnetic field strengths found in a field of view at
$l = 153\dg$, $0.5\dg < b < 18\dg$ by \cite{skh07b} and at high
Galactic latitudes $b = 70\dg$ \citep{dkh06} probably reflect
smaller-scale magnetic field fluctuations and not the large-scale
field.

\section{Magnetic fields in the entire disk+halo system}
\label{s:diskhalo}

Most models in Table~\ref{t:bfldmodels} do not only discuss the
Galactic disk or halo but simultaneously fit both the disk and the
halo, allowing for different configurations in disk and halo magnetic
fields.  \cite{jfw09} tried to unify models of the Galactic disk and
halo, using a number of different models from the literature. They
concluded that the magnetic field structure in the Galactic disk and
halo were different and cannot be captured by scaled-up versions of
the same magnetic field configuration. Their best-fit model is a
disk-even halo-odd (DEHO) field, which was shown to be theoretically
possible if one attributes an important role to the Galactic wind in
the dynamo process \citep{msb10}. The conclusion of a best-fit DEHO
field was also reached by \cite{srw08}.  These authors fitted 22.8~GHz
synchrotron data from the Wilkinson Microwave Anisotropy Probe
\citep[WMAP,][]{hnb07} and rotation measures from 1090 extragalactic
point sources to their models and argued that none of the available
models were a good fit to the data. However, they could conclude that
a disk magnetic field which was symmetric with respect to the Galactic
plane was strongly favored. For the halo, a toroidal field which is
anti-symmetric with respect to the plane was preferred. Indeed, this
anti-symmetric structure in the rotation measure sky with respect to
the Galactic Center (``butterfly pattern''), only existent in the
inner Galaxy, was already noticed decades ago \citep{sk80}. This has
been interpreted as an $A0$ dynamo \citep{hml06}, i.e.\ a dynamo causing
an $A0$ field configuration (see Section~\ref{s:lsvss}), but has also been
attributed to local structure \citep{wfl10,sts11}. $A0$ dynamo models
are also strongly inconsistent with modeling comparing near-infrared
starlight polarization measurements, based on discrepancies in the
Galactic disk. However, adding an even disk-component to these models
(DEHO field) makes them inconsistent with observed degrees of
polarization of near-infrared starlight \citep{pcp12}.

\cite{jlb10} included an ordered (anisotropic random) component to the
regular (coherent) and random Galactic magnetic field components in
their model.  Due to the large number of free parameters (22), some of
which are degenerate, they choose to constrain some parameters using
one observational data set only, and keeping these fixed while
constraining other parameters. Due to these degeneracies and the large
number of unknowns, the authors caution to not attach too much value
to the absolute numbers they find for field strengths. They do argue
that their ratio of the three field components
regular:random:anisotropic random, of 1:5:3, is relatively robust. So
they conclude that the anisotropic random field is stronger than the
regular field component.  In \cite{jbl11}, these authors use a more
realistic cosmic ray distribution in the Galaxy and find that the
random component is even larger with respect to the coherent
component.

At the moment, the latest all-inclusive modeling attempt is presented
in \cite{jf12a, jf12b}. \cite{jf12a} include two new components for
the magnetic field: a vertical, out-of-plane component similar to the
X-shaped fields seen in external galaxies \citep[e.g.,][]{hh12}; and a
contribution by anisotropic random magnetic fields. The latter is
degenerate with an increased intensity of cosmic ray electrons over
the usually quoted values \citep{smp07}, but generally comparable in
strength to the regular field. A random component for the magnetic
field is added in \cite{jf12b}, which is allowed to vary in strength
in 8 spiral regions. This complex magnetic field model now has 36 free
parameters, making it exceedingly difficult to be confident that the
true minimum in 36-dimensional parameter space has been found.

The set of papers which fit magnetic field models to radio
polarization data at high frequencies ($\geq 22$~GHz WMAP data)
(\cite{rrb10,fma11}) tend to find a higher pitch angle than rotation
measure studies of $\sim 24\dg-30\dg$. Planck all-sky maps will
provide additional observational constraints, which simulations show
suggest the same high pitch angles \citep{fmj12}.  A non-negligible
vertical magnetic field component is needed in these models as well,
currently at odds with conclusions from rotation measure analyses
(Section~\ref{s:halo}).

It may be possible in the near future to derive Galactic magnetic
field structure from observations of arrival directions of Ultra-High
Energy Cosmic Rays (UHECRs), but currently the sources and composition
of UHECRs are too uncertain to constrain any Galactic magnetic field
models \citep{ghm09}.

\section{Summary and conclusions}
\label{s:conc}

In this Section, I present a short summary of observational
knowledge of magnetic fields in the Milky Way, neglecting all
subtleties discussed above. I will also try to draw some conclusions.

The strength of the magnetic field in the Solar neighborhood is fairly
well determined. The regular, large-scale component $B_{reg}\approx
2~\mu$G, while the total magnetic field is $B_{tot}\approx
6~\mu$G. Estimates of the isotropic random magnetic field from
magnetic field modeling of $B_{ran}\approx 3-4~\mu$G suggest that
there exists also an anisotropic random field component of comparable
strength to the random component. The magnetic field strength
increases towards the inner Galaxy, and is independent of density for
the diffuse interstellar gas.

The magnetic field direction in the Galactic disk most likely roughly
follows the spiral arms. This is not always the case, since pitch
angle estimates from modeling still vary, and there are concrete
indications at several locations in the Galactic disk that the
magnetic field direction does not coincide with the stellar or gaseous
arms. The disk magnetic field is symmetric (even) with respect to the
Galactic plane.

There is one large-scale magnetic reversal close to the Sun towards
the inner Galaxy, but the existence and location(s) of more reversals
is still under debate. The studies that rely mostly or totally on
pulsar data indicate magnetic fields with more intermediate-scale
structure (reversals) than studies (also) including extragalactic
source RMs. This difference is likely due to the intrinsic differences
in the data: extragalactic source RMs are averaged over the entire
line of sight through the Galaxy and often over a patch of the plane
of the sky as well, which washes out smaller scale structure
partially. Fitting pulsar data would retrieve this smaller scale
structure. However, pulsars with known RMs are concentrated in a few
kpc from the Sun and their distances can be quite
uncertain. Therefore, it is quite possible that what is interpreted as
reversals along spiral arms are actually other intermediate scale
structures caused by e.g.\ superbubbles.

The pitch angle of the magnetic field is roughly $p \approx -5\dg$ to
$-15\dg$, depending on tracer (rotation measure, starlight
polarization, gas, CO, etc). A notable exception is modeling of
high-frequency (i.e.\ Planck/WMAP frequencies and higher) synchrotron
emission, studies of which consistently show higher pitch angles of
$p\approx -25\dg$ to $-30\dg$. The random magnetic field component
shows a turbulent power spectrum with an outer scale of turbulence
that is a few parsecs in the disk, possibly only the spiral arms, and
up to $\sim100$~pc in the Galactic halo.

The magnetic field strength in the gaseous halo, or thick disk, is
comparable to that in the disk, with an uncertainty of a factor
$2-3$. The scale height is many kiloparsecs ($\sim5-6$~kpc), possibly
smaller for the regular field component. There is a pronounced
north-south asymmetry across the Galactic disk: the magnetic field
variance is higher in the south. There is a small large-scale vertical
magnetic field component towards the south $B_z\approx 0.3~\mu$G,
while a small vertical magnetic field component towards the north
could be attributed to the North Polar Spur.

The complete disk-halo system has been extensively modeled in the past
decade, using a wide variety of observational tracers, magnetic field
configurations and components, thermal and cosmic ray density models,
and input and output parameters. One property that all models share is
that none of them gives a satisfactory fit to all the data. This is
not surprising, seeing the immense complexity of the magnetic and
gaseous structures observed in the Milky Way. Large loops of radio
emission such as the North Polar Spur or Loop I to IV \citep{b71} show
influence of magnetic fields \citep{,s73,sts11}, created by supernovae
blowing bubbles in the ionized interstellar gas, dragging the magnetic
field with them. The named Loops are giant structures in the sky
because they are located very close to the Sun and are therefore
conspicuous on the sky. However, hundreds or even thousands more of
these structures should exist in the rest of the Milky Way, all
affecting the large-scale structure of the magnetic field. These and
other local structures are virtually impossible to include in modeling
and therefore often omitted. This is especially clear towards the
Galactic anti-center, where the regular magnetic field is directed
almost perpendicular to the line of sight and therefore has a
negligible RM contribution in this direction. As \cite{ptk11} note,
any regular magnetic field model with a small pitch angle severely
underestimates the amount of RM fluctuations observed in this
direction. These local structures, combined with the
location-dependent turbulent nature of the magneto-ionized medium,
make this modeling a daunting enterprise.

The variety in conclusions from Galactic magnetic field models using
different tracers and methods, indicates a large role of small-scale
position-dependent turbulence, discrete structures, significant
changes in pitch angle along a spiral arm, or -- most likely -- all of
these.  Variable pitch angles are also suggested by simulations of
density waves including magnetic fields \cite{gc04}. This explanation
does make it more plausible why a ring-like magnetic field model gives
fit results of comparable quality as the spiral arm models, or why
deviations of magnetic field directions from gaseous and stellar pitch
angles are found.

\section{Epilogue}
\label{s:future}

A number of recent technological and computational developments make a
large expansion in parameter space related to studies of cosmic 
magnetic fields possible: Phased Array Feeds allow deep surveys of
large parts of the sky in reasonable observing times; low frequency
polarimetry is becoming possible thanks to sufficient computer power
and technological expertise to build software telescopes, and finally
large-scale galactic (but also extragalactic, intracluster) magnetic
fields can be probed in (almost) three dimensions using Rotation
Measure Synthesis \citep{bd05}.

This has sparked renewed interest in the field of cosmic magnetism, as
evidenced by the {\em Cosmic Magnetism Key Science Project}
\citep[MKSP,][]{abb12} for the LOw-Frequency ARray LOFAR; the {\em
  Polarisation Sky Survey of the Universe's Magnetism}
\citep[POSSUM,][]{glt10} for the Australian Square Kilometre Array
Pathfinder (ASKAP), and cosmic magnetism studies as part of the WODAN
project \citep{rab11} using the APERTIF Phased Array Feeds on the
Westerbork Synthesis Radio Telescope (WSRT). These are all exciting
innovative telescopes and/or instruments currently under construction.
For details on magnetism studies with LOFAR, SKA, Planck and ALMA see
the Chapters by R. Beck and W. Vlemmings in this Volume. I will
discuss other important future and ongoing initiatives below.

\subsection{Galactic magnetism with existing instrumentation}

A number of large radio polarimetric surveys have recently been done
or are in progress, with the aim of studying the magnetized
 ISM of the Milky Way, at a variety of frequencies.

 Several surveys with the ALFA seven feed array on the Arecibo
 telescope are being performed, among which the Galactic ALFA
 Continuum Transit Survey \citep[GALFACTS,][]{ts10}. GALFACTS will
 survey the whole Arecibo sky (declinations $-1.33\dg < \delta <
 38.03\dg$) in the frequency range $1225-1525$~MHz down to a
 sensitivity of $90~\mu$Jy. Its main science goals are exploration of
 the Milky Way's magnetic field and the properties of the magnetized
 ISM. Observations have been progressing for four years and will be
 completed in 2013.

The lower Faraday rotation (and therefore more distant polarization
horizon) at higher frequencies was the reason for the 6-cm Sino-German
survey of the Galactic plane ($10\dg<l<230\dg$, $|b| < 5\dg$) with the
Urumqi 25-m single dish \citep{shr07, grh10, srh11, xhr11}. This
survey is mostly focused on the detection of discrete magnetized
objects such as H~{\sc ii} regions, supernova remnants and Faraday
screens. At even higher frequencies of 5~GHz, the C-Band All-Sky
Survey (C-BASS) will provide an all-sky polarimetric survey. Although
its main science goal is providing characterization of foregrounds for
Cosmic Microwave Background (CMB) polarization studies, it will also
explore Galactic magnetic fields. Data acquisition is ongoing.

The S-Band Polarization All-Sky Survey (S-PASS) is a radio
polarimetric study of the entire southern sky at 2307~MHz in a 184~MHz
bandwidth, performed with the Parkes 64m single dish telescope with a
polarization sensitivity better than 1~mJy/beam. The science goals of
the survey are two-fold: characterizing polarized foregrounds for
measurements of the B-mode of CMB 
Polarization, and exploration of Galactic magnetic fields. The survey
observations are completed and first science results are being
published \citep{ccs13,cbs13,sgc13}. The Southern Twenty-centimeter All-sky
Polarization Survey (STAPS) was observed commensally with S-PASS and
data processing is ongoing.

The largest ongoing project to map Galactic magnetism using existing
instrumentation is the Global Magneto-Ionic Medium Survey
\citep[GMIMS,][]{wlc09}. This project consists of a series of
polarimetric surveys in the northern and southern hemispheres, from
$\sim300$~MHz to $\sim1800$~MHz. Data acquisition for the southern-sky
survey spanning $287-870$~MHz with the Parkes telescope is completed
and data processing in progress, while the STAPS survey described above
will function as the high-band ($1300-1800$~MHz) southern-sky survey
for GMIMS. For the high-band survey in the north ($1277-1740$~MHz),
performed with the DRAO 26-m single dish \citep{wlh10}, observations
have finished and data reduction is nearing completion, with first science
results discussed in \cite{wfl10}. Options for observing the
remaining GMIMS surveys are being considered.

With an angular resolution of $30-60^{\prime}$ and a frequency
resolution of at least 1~MHz, GMIMS will provide the first
spectro-polarimetric data set of the large-scale polarized emission
over the entire sky, observed with single-dish telescopes. The broad
frequency coverage is of great importance for high resolution and
broad sensitivity of Rotation Measure Synthesis. Therefore, the
combined surveys with a 1500~MHz bandwidth will give unprecedented
maps of Faraday depth over the whole sky, revolutionizing studies of
the magneto-ionized ISM in the Galaxy using this
method.

\subsection{Galactic magnetism with next-generation instrumentation}

The WSRT is being upgraded with phased array feeds named APERture Tile
In Focus \citep[APERTIF,][]{ovc10} with a 300~MHz bandwidth in the
range of 1.0~GHz to 1.7~GHz. This upgrade will increase Westerbork's
field of view with a factor 25 to about 8~square degrees, making it a
wonderful survey instrument.

One of the key surveys to be performed with APERTIF is the Westerbork
Observations of the Deep APERTIF Northern-Sky
\citep[WODAN,][]{rab11}. WODAN aims to image the whole northern sky
down to $10~\mu$Jy rms with a broad bandwidth around 1400~MHz, and
part of the sky a factor two deeper. It is mostly geared towards
cosmology and other extragalactic science, with an observational aim
to detect 30~million radio sources including 100,000~clusters,
10~million starbursting galaxies at $z>1$ and virtually all radio loud
AGN in the Universe. However, many of these sources will emit
polarized emission at this wavelength, which will be Faraday rotated
by the Galactic magnetized ISM. This will provide an
observational data set for Galactic magnetism studies far surpassing
the currently available NVSS rotation measure data base
\citep{tss09}. 

Similarly to WODAN in the northern sky, the southern sky will be
surveyed by several projects on the Australian SKA Pathfinder
\citep[ASKAP,][]{jtb08}. WODAN's sister survey is called Evolutionary
Map of the Universe \citep[EMU,][]{n10}, but the data obtained is
shared between EMU and a project dedicated to cosmic magnetism, named
Polarization Sky Survey of the Universe's Magnetism
\citep[POSSUM,][]{glt10}. POSSUM aims to measure the Faraday rotation
of 3~million extragalactic radio sources over 30,000~square degrees,
which will allow major steps in characterizing the large-scale and
turbulent components of the Galactic magnetic field, but also test
(dynamo) theories for the origin and evolution of the Milky Way's
magnetic field.

Finally, the Murchison Widefield Array \citep[MWA,][]{wbb11}, under
development in Western Australia at the moment is a low-frequency
radio interferometer at $80-300$~MHz - analogous to LOFAR in the north
but smaller; however, with an excellent uv-coverage on small
baselines. Although its main science goals are the Epoch of
Reionization, solar and ionospheric science and transients, it can
also be used to provide detailed rotation measure synthesis maps of
low-magnetic-field areas in the southern sky.

\section*{Acknowledgments}

The author wants to express her sincere thanks to Katia Ferri\`ere and
George Heald for critically reading and commenting on the manuscript,
to Philip Kronberg for discussion about details of spiral arms pitch
angles, to Jo-Anne Brown, Tess Jaffe, Aris Noutsos, and Cameron van
Eck for kindly giving permission to use their figures and useful
comments and discussion. This work is part of the research programme
639.042.915, which is (partly) financed by the Netherlands
Organisation for Scientific Research (NWO).

%
%
 \bibliographystyle{science-online,unsrtnat}
 \bibliography{}
\newcommand{\aap}{A\&A}
\newcommand{\apj}{ApJ}
\newcommand{\mnras}{MNRAS}
\newcommand{\jcap}{JCAP}
\newcommand{\pasa}{PASA}
\newcommand{\ssr}{SSRev}
\newcommand{\apjs}{ApJS}
\newcommand{\apjl}{ApJL}
\newcommand{\aj}{AJ}
\newcommand{\araa}{ARA\&A}
\newcommand{\apss}{Ap\&SS}
\newcommand{\nat}{Nature}

\end{document}